\documentstyle[aps,prd,eqsecnum,twocolumn,epsf]{revtex}
\newcommand{\beq}{\begin{equation}}
\newcommand{\beqn}{\begin{eqnarray}}
\newcommand{\eeq}{\end{equation}}
\newcommand{\eeqn}{\end{eqnarray}}

\newcommand{\gsim}{\mbox{\raisebox{-1.ex}{$\stackrel
     {\textstyle>}{\textstyle\sim}$}}}
\newcommand{\lsim}{\mbox{\raisebox{-1.ex}{$\stackrel
     {\textstyle<}{\textstyle \sim}$}}}
\newcommand{\square}{\kern1pt\vbox{\hrule height
1.2pt\hbox{\vrule width 1.2pt\hskip 3pt
   \vbox{\vskip 6pt}\hskip 3pt\vrule width 0.6pt}\hrule
height 0.6pt}\kern1pt}

\begin{document}

\draft
\twocolumn[\hsize\textwidth\columnwidth\hsize\csname
@twocolumnfalse\endcsname

\title{Brane preheating} 
\author{Shinji Tsujikawa$^1$, Kei-ichi Maeda$^{1,2}$ and Shuntaro 
Mizuno$^1$ \\~} 
\address{$^1$Department of Physics, Waseda University, Okubo 3-4-1,
Shinjuku, Tokyo 169-8555, Japan\\[-1em]~}
\address{$^2$ Advanced Research Institute for Science and Engineering, 
Waseda University, Shinjuku, Tokyo 169-8555, Japan\\[-.7em]~}

\date{\today} 
\maketitle
\begin{abstract}
We study brane-world preheating in massive chaotic inflationary scenario 
where scalar fields are confined  on the 3-brane.  Assuming that quadratic 
contribution in energy densities  dominates the Hubble expansion rate during preheating,
the  amplitude of inflaton decreases slowly relative to the standard 
dust-dominated case. This leads to an efficient production of $\chi$ particles 
via nonperturbative decay of inflaton even if its coupling is of order 
$g=10^{-5}$.  We also discuss massive particle creation  heavier  than
inflaton, which may play important roles for the baryo-  and
lepto-genesis scenarios.
\end{abstract}
\vskip 0.5pc 
\centerline{WUAP-00/29}
\vskip 1pc
]

%%%%%%%%%%%%%%%%%%%%%%%%%%%%%%%%%%%%%%%%%%%%%%%%%%
%                                                %
\section{Introduction}                           %
%                                                %
%%%%%%%%%%%%%%%%%%%%%%%%%%%%%%%%%%%%%%%%%%%%%%%%%%

Recent studies of superstring or M-theory   predict that our universe may 
consist of a brane world embedded in a higher-dimensional spacetime 
\cite{braneori,HW,Anto}.  
In particular,  it was shown that 
the 10-dimensional $E_8 \times E_8$ heterotic 
string theory, which is a strong candidate to describe 
our real world, is equivalent to an 11-dimensional M-theory 
compactified to ${\bf M}^{10} \times {\bf S}^1/Z_2$\cite{HW}.
Then the 10-dimensional spacetime is 
expected to be compactified into ${\bf M}^{4}\times {\bf CY}^{6}$.  Therein 
the standard model particles are confined on  the 3-dimensional brane.  On 
the other hand, gravitons propagate in the full spacetime.

In conventional Kaluza-Klein theories, extra dimensions are 
compactified on some manifolds in order to obtain 4-dimensional effective 
gravity theories.  In contrast, Randall and Sundrum \cite{RS} recently 
proposed a new type of compactification, in which extra dimension is not 
compact but gravity is effectively 3-dimensional.  Their model indicates 
that we are living on the 3-brane with positive tension embedded in 
5-dimensional anti de-Sitter bulk spacetime.  This idea invokes a 
particular interest for the evolution of the early universe, which may 
exhibit some deviations from the standard cosmology by the modifications of 
4-dimensional Einstein equations on the brane \cite{brane_cos,SMS}.

In this respect, cosmological inflation was investigated by several 
authors\cite{MWBH,brane_inflation}
in the context of the brane-world scenario where scalar fields are confined 
on the 3-brane. 
The effect of the brane in higher dimensions induces  extra terms in 
4-dimensional effective gravity  especially for high energies, e.g., a
quadratic term of energy density, 
which leads to the larger amount of inflation relative to  standard 
inflationary scenarios.  
 
If the modifications due to extra dimensions persist in the post inflationary
phase, this may nontrivially affect on the process of 
reheating.  In the early stage of reheating, particles coupled to inflaton are 
efficiently created by the nonperturbative and  
nonequilibrium process which is 
called preheating\cite{TB,KLS,KT,Boy}.  Since the particle production during 
preheating is sensitive to the background evolution\cite{KLS}, it is 
expected that the standard picture of preheating will be altered due to the 
modifications in Einstein equations on the brane.  In fact we will show 
that the brane-world preheating broadens the parameter ranges of coupling 
constants where particles are efficiently produced.

%%%%%%%%%%%%%%%%%%%%%%%%%%%%%%%%%%%
 \section{Brane inflation }
%%%%%%%%%%%%%%%%%%%%%%%%%%%%%%%%%%%
In the brane world scenario by Randall-Sundrum\cite{RS}, 
where gravity as well as 
matter field are effectively  confined on the 3-brane 
in the 5-dimensional spacetime,
the 5-dimensional Einstein  equations can be recast in the following 
gravitational equations on the brane \cite{SMS}: 
\beqn
^{(4)}G_{\mu\nu}=-\Lambda_4 g_{\mu\nu}+\frac{8\pi}{M_{\rm pl}^2} 
T_{\mu\nu}+\frac{1}{M_5^6} \pi_{\mu\nu}-E_{\mu\nu},
\label{geq}
\eeqn 
where $T_{\mu\nu}$ and $\pi_{\mu\nu}$ represent the energy-momentum 
tensor on the brane and the quadratic term in $T_{\mu\nu}$, respectively.  
$E_{\mu\nu}$ is a part of 5-dimensional Weyl tensor, which carries the 
information in the bulk.  Note that the four- and five-dimensional 
gravitational constants, $M_{\rm pl}$ and $M_5$, are related with the 
3-brane tension, $\lambda$, as $ \lambda=48\pi {M_5^6}/{M_{\rm pl}^2}.  
$\footnote{Our notation of the 5-dimensional Planck mass, $M_5$, does not 
include the $\sqrt{8\pi}$ factor, which is different from the definition in 
Ref.~\cite{MWBH}.} 
Here the 4-dimensional cosmological constant  
 $\Lambda_4$ is assumed to be zero. 

Let us analyze the gravitational equation (\ref{geq}) in cosmological 
contexts.  Adopting a flat Friedmann-Robertson-Walker (FRW) metric as a 
background spacetime on the brane, we find the following Friedmann 
equation\cite{brane_cos,SMS,MWBH} 
\beqn
H^2 \equiv \left(\frac{\dot{a}}{a}\right)^2=\frac{8\pi}{3M_{\rm pl}^2} 
\rho \left(1+\frac{\rho}{2\lambda}\right),
\label{hubble}
\eeqn 
where $a$ and $\rho$ are the scale factor and the energy density of the matter
on the brane, respectively.  We have ignored the so-called  ``dark" radiation 
$E_{\mu\nu}$, which rapidly decreases as $\sim a^{-4}$ during 
inflation even if it exists.

At high energies the $\rho^2$ term is expected to play important roles 
to determine the evolution of the Universe.  Hereafter we will assume that 
the quadratic term dominates during inflation and preheating, i.e., the 
condition, $\rho~\gsim~2\lambda$, holds until the end of preheating.
The inflaton field confined on the brane satisfies 
the Klein-Gordon equation, 
\beqn
\ddot{\phi}+3H\dot{\phi}+V'(\phi)=0.
\label{phi}
\eeqn 
Note that during inflation the quadratic contribution in Eq.~(\ref{hubble}) 
increases the Hubble expansion rate,  which makes the evolution of inflaton 
slower by Eq.~(\ref{phi}).
Another Einstein equation in the quadratic-term dominant stage yields
\beqn
\frac{\ddot{a}}{a}=\frac{\pi}{3M_{\rm pl}^2\lambda} 
\left(\dot{\phi}^2+2V \right) \left(2V-5\dot{\phi}^2\right).
\label{ddota}
\eeqn 
This implies that inflation ends around $2V \simeq 5\dot{\phi}^2$.
In the massive chaotic inflationary scenario with a potential, 
$V(\phi)=\frac12 m^2\phi^2$, this condition gives 
\beqn
\phi_F \simeq 3\left(\frac{M_5}{m}\right)^{1/2} M_5,
\label{phiend}
\eeqn 
where we used the slow-roll condition, 
$\dot{\phi}\simeq -m^2\phi/3H \simeq -4M_5^3/\phi$ in Eq.~(\ref{phi}).

The inflaton mass, $m$, should be constrained by comparing the amplitude of 
density perturbations produced during inflation with the CMB anisotropy.  
This is typically done by evaluating the curvature perturbation, $\zeta$, 
around the 55 e-foldings before the end of inflation, if $\zeta$ is 
conserved on large scales.  It was shown in Ref.~\cite{Wands} that the 
conservation of $\zeta$ holds for adiabatic perturbations irrespective of 
the form of gravitational equations by considering the local conservation 
of the energy-momentum tensor.  In the presence of a second scalar field, 
$\chi$, curvature perturbations include isocurvature (entropy) 
perturbations\cite{Wands,GWBM}, which provides the source of nonadiabatic 
pressure perturbations.  However, as long as $\chi$ does not exhibit strong 
growth during inflation as in the nonminimally coupled case\cite{TY}, 
$\zeta$ remains conserved even in the multi-field case\cite{multi}.  In 
fact, in the massive chaotic inflationary scenario with the interaction 
$\frac12 g^2\phi^2\chi^2$, $\chi$ is suppressed during inflation for large 
couplings $g$ required for the $\chi$ particle production during 
preheating\cite{sup}, which ensures the conservation of $\zeta$ on large 
scales.  Although we will use this result here, the Weyl tensor $E_{\mu\nu} 
$ will bring the 5-dimensional information into the brane, which may alter 
the result \cite{brane_perturbation1}.  More careful analysis will be 
required regarding the evolution of $\zeta$
during inflation\cite{brane_perturbation2}.

%%%%%%%%%%%%%%%%%%%%%%
\begin{figure}
\epsfxsize = 3.5in
\epsffile{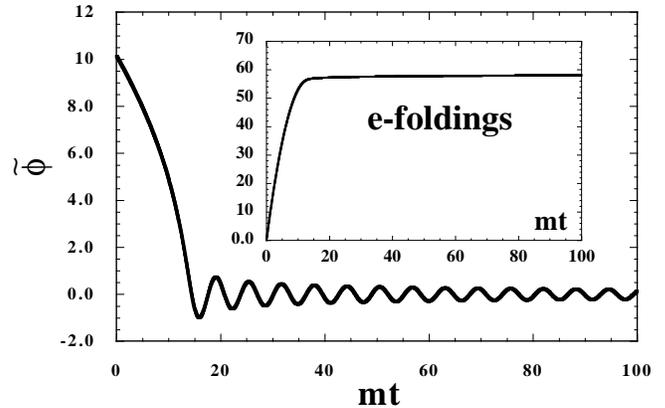} 
\caption{The evolution of the inflaton condensate, 
$\tilde{\phi}=(m/M_5)^{1/2}\phi/M_5$, during inflation and reheating.
We start integrating with the initial value in Eq.~(\ref{phicobe}). Inflation 
ends around $\tilde{\phi} \simeq 3$ as estimated by Eq.~(\ref{phiend}). 
The system enters the reheating stage around $\tilde{\phi} \simeq 1$.
{\bf Inset:} The evolution of the number of e-foldings, 
$N=\int_{t_*}^{t_F} H dt$. We find that 55 e-foldings are achieved
during inflation.}
\label{Fig1}
\end{figure}
%%%%%%%%%%%%%%%%%%%%%%

Calculating the number of e-foldings: $N \equiv \int_{t_*}^{t_F} H dt$, and 
setting $N=55$, we obtain the value of inflaton when scales relevant for the 
CMB anisotropy crossed outside the Hubble radius,
\beqn
\phi_* \simeq 10\left(\frac{M_5}{m}\right)^{1/2} M_5.
\label{phicobe}
\eeqn 
The spectra of density perturbations at horizon re-entry can be evaluated 
as  $\delta_H=(2/5)|\zeta|_{k=aH}=(2/5)H\delta\phi/\dot{\phi}|_{k=aH}$,
where the subscript, $k=aH$, denotes the values at first horizon crossing.
Making use of the relation $\delta\phi \simeq H/2\pi$ at first horizon 
crossing\footnote[2]{While the formula used here is based on the assumption of 
the Bunch-Davies vacuum,  the present
expansion law is different from the de Sitter one, which may change the 
expression of $\delta_H$.}, we find 
\beqn
\delta_H \simeq \frac{1}{\left(24\sqrt{5\pi}\right)^2} 
\left(\frac{m}{M_5}\right)^4 \left(\frac{\phi_*}{M_5}\right)^{5}.
\label{deltaH}
\eeqn 
The COBE normalization gives $\delta_H \simeq 2 \times 10^{-5}$,
which constrains the mass of inflaton as 
\beqn
m/M_5 \simeq 1.4 \times 10^{-4}.
\label{mass}
\eeqn 
The ratio of 
Eq.~(\ref{mass}) is crucial to determine the efficiency of parametric 
resonance during preheating.  We should caution again that the result of 
Eq.~(\ref{mass}) could be modified by a careful study of the quantum theory 
in the brane-world inflation and the evolution of cosmological perturbations 
in the 5-dimensional Einstein equations 
\cite{brane_perturbation1,brane_perturbation2}.

We have numerically 
solved the background equations (\ref{hubble}) and (\ref{phi}) with the 
initial value (\ref{phicobe}), and confirmed that analytic estimations are 
in good agreement with numerical results (see Fig.~1).

%%%%%%%%%%%%%%%%%%%%%%%%%%%%%%%%%%%%%%%%%%%%%%%%%%
 \section{Brane preheating}
%%%%%%%%%%%%%%%%%%%%%%%%%%%%%%%%%

Let us investigate a two-field model where the second scalar field, $\chi$, is 
coupled to massive inflaton, $\phi$, with a four-leg interaction, $\frac12 
g^2\phi^2\chi^2$: 
\beqn
V(\phi,\chi)=\frac12 m^2\phi^2+\frac12 g^2\phi^2\chi^2
+\frac12 m_{\chi}^2\chi^2,
\label{effpotential}
\eeqn 
where $m_{\chi}$ is a mass of the $\chi$ field.
We consider perturbations of scalar fields on the 3-brane in the 
flat FRW background and implement backreaction effects of 
produced particles at second order \cite{KLS}. 

Decomposing the scalar fields into homogeneous parts and 
fluctuations as $\phi(t,{\bf x}) \to \phi(t)+\delta \phi(t, {\bf x})$
and $\chi(t,{\bf x}) \to \chi(t)+\delta \chi(t, {\bf x})$, the 
equation for the homogeneous part of inflaton can be written as 
\beqn
\ddot{\phi}+3H\dot{\phi}+\left(m^2+g^2
\langle\chi^2\rangle \right)\phi=0,
\label{phit}
\eeqn 
where $\langle\chi^2\rangle$ is the expectation value of $\chi^2$, which 
works as backreaction at the final stage of preheating.  Neglecting the 
contribution of inflaton fluctuations which can be negligible in the 
present model unless rescattering effects are taken into account, each 
Fourier mode of the $\delta\chi_k$ fluctuation satisfies 
\beqn
\delta\ddot{\chi}_k+3H\delta\dot{\chi}_k+
\left[\frac{k^2}{a^2}+g^2\phi^2(t)+m_{\chi}^2 \right]
\delta\chi_k=0.
\label{deltachi}
\eeqn 
The oscillation of the inflaton condensate leads to parametric
amplification of the $\delta\chi_k$ fluctuation, whose variance is defined by 
\beqn
\langle \delta\chi^2 \rangle = \frac{1}{2\pi^2}
\int k^2 |\delta\chi_k^2| dk.
\label{varaince}
\eeqn 
Note that the equation for $\chi(t)$ is obtained by substituting 
$k=0$ in Eq.~(\ref{deltachi}), whose evolution is similar to 
the long wave $\delta\chi_k$ modes.

Assuming that the linear term of $\rho$ in Eq.~(\ref{hubble}) can 
be neglected relative to the quadratic term during the whole stage of 
preheating, the Hubble parameter satisfies 
\beqn
H=\frac{1}{6M_5^3} \Bigl[ \frac12 \dot{\phi}^2 &+& \frac12 
m^2\phi^2+\frac12 \dot{\chi}^2+ \frac12 m_{\chi}^2\chi^2+\frac12 
g^2\phi^2\chi^2 \nonumber \\
&+& \rho_{\phi}+\rho_{\chi} \Bigr],
\label{hubble2}
\eeqn 
where $\rho_{\phi}$ and $\rho_{\chi}$ are the energy densities
of produced particles.  In the stage where contributions of $\chi$ 
are negligible in Eqs.~(\ref{phit}) and (\ref{hubble2}),
making use of the time averaged relation, 
$\langle\frac12\dot{\phi}^2\rangle _T=\langle V(\phi) \rangle_T$,
gives the following approximate solutions for the scale factor 
and the background inflaton field:
\beqn
a = a_i \left(\frac{t}{t_i} \right)^{1/3},~~~  
\phi(t) = \Phi(t) \sin mt,
\label{aapprox}
\eeqn 
where the amplitude, $\Phi(t)$, is
\beqn 
\Phi(t)=2\left(\frac{M_5}{m}\right)^{3/2} 
\sqrt{\frac{m}{t}}.
\label{phiapprox}
\eeqn 
This evolution is different from the standard results
in the dust-dominated preheating scenario,
$a \propto t^{2/3}$ and $\Phi(t) \propto t^{-2}$ \cite{KLS}. 
Particle creations in an expanding universe are strongly affected 
by the change of background evolution.

Introducing a new variable, $\delta X_k=a^{3/2}\delta\chi_k$,
and neglecting a curvature term in the frequency of $\delta X_k$
which is unimportant during reheating, we find that 
Eq.~(\ref{deltachi}) is reduced to the Mathieu equation
\beqn 
\frac{d^2}{dz^2} \delta X_k+
\left(A_k-2q \cos2z \right)\delta X_k=0,
\label{mathieu}
\eeqn 
where $z=mt$ and 
\beqn 
A_k &=& 2q+\frac{k^2}{m^2a^2}+\frac{m_{\chi}^2}
{m^2}, \nonumber \\
q &=& \frac{g^2\Phi^2(t)}{4m^2}
=g^2\left(\frac{M_5}{m}\right)^3 \frac{1}{z}.
\label{rparameter}
\eeqn 
The efficiency of resonance strongly depends upon the 
initial value of $q$ ($=q_i$).
In the standard preheating scenario where $q$ decreases as 
$q \sim t^{-2}$, we require the coupling, $g~\gsim~10^{-4}$ 
which corresponds to $q_i~\gsim~100$, for the relevant 
growth of $\delta\chi_k$ fluctuations\cite{KLS}.

In contrast, in the brane-world preheating, $q$ evolves more slowly as $q 
\sim t^{-1}$, which indicates that large $q_i$ much greater than unity is 
not required for $\chi$ particle productions to occur.  This has some 
analogies with the nonminimally coupled $\chi$ field in the $R^2$ inflation 
model where slow decrease of the oscillating scalar curvature ($R \propto 
t^{-1}$) leads to strong preheating\cite{nonminimal}.  The ratio $m/M_5$ is 
an important factor to determine the strength of resonance.  Adopting the 
value of Eq.~(\ref{mass}) which comes from the CMB constraints, one obtains 
\beqn 
q \simeq 0.36\left(10^6g\right)^2 \frac{1}{z}.
\label{qest}
\eeqn 
The coherent oscillation of inflaton turns on around $z \simeq \pi/2$, 
which yields the relation, $q_i \simeq (10^6g)^2/5$.  
When $m_{\chi}=0$, we find numerically that $\delta\chi_k$ fluctuations 
exhibit exponential increase for $g~\gsim~5 \times 10^{-6}$, corresponding 
to $q_i~\gsim~5$.

Let us consider concrete cases with a massless $\chi$ field.  When $g=5 \times 
10^{-6}$, the variance $\langle\delta\chi^2\rangle$ increases at the 
initial stage, but its growth stops around $mt \simeq 35$ (see Fig.~2).  
Although $\chi$ particles are created while $\delta\chi_k$ fluctuations 
pass through instability bands with the decrease of $q$ due to cosmic 
expansion, no creation occurs once $q$ drops down to $1/4 \sim 1/3$, since 
no instability bands exist there\cite{KLS}.  Actually, substituting $z=35$ 
for Eq.~(\ref{qest}) gives the value, $q \simeq 0.26$, which coincides well 
with the above estimation.

%%%%%%%%%%%%%%%%%%%%%%
\begin{figure}
\epsfxsize = 3.5in
\epsffile{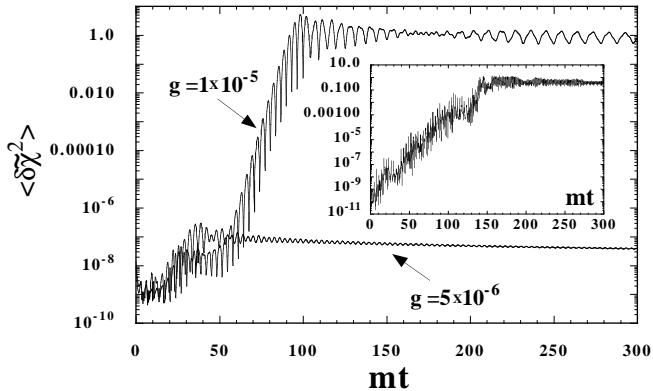} 
\caption{The evolution of the variance, 
$\langle\delta\tilde{\chi}^2\rangle
\equiv \langle\delta\chi^2\rangle/M_5^2$,
 during preheating for 
$g=5 \times 10^{-6}$ and $g=1 \times 10^{-5}$ with $m_{\chi}=0$.
While resonance soon terminates at the initial stage for $g=5 \times 10^{-6}$,
$\chi$ particles are sufficiently produced for $g=1 \times 10^{-5}$.
{\bf Inset}: $\langle\delta\tilde{\chi}^2\rangle$ vs $t$ for the  case of 
$g=2 \times 10^{-3}, m_{\chi}=10m$.
The $\chi$ particle whose mass is greater than the inflaton mass, $m$,
can be sufficiently created for $g>2 \times 10^{-5}$. }
\label{Fig2}
\end{figure}
%%%%%%%%%%%%%%%%%%%%%%

With the increase of $g$, the duration during which the $\chi$ field
stays in instability bands gets longer.
In Fig.~2 we plot the evolution of $\langle\delta\chi^2\rangle$ for
$g=10^{-5}$.  We find that the field reaches a short plateau
around $25~\lsim~mt~\lsim~50$ after some growth in the initial stage.
This corresponds to the stability region around $1~\lsim~q~\lsim~2$, after 
which $\langle\delta\chi^2\rangle$ begins to grow again since the field 
passes through the first instability band around $1/3~\lsim~q~\lsim~1$.  In 
this case the backreaction effect of created $\chi$ particle is marginally 
important at the final stage of preheating.

When $g~\gsim~10^{-5}$, the backreaction effect plays an important 
role to shut off the resonance.  While the growth rate of 
$\delta\chi_k$ fluctuations gets larger with $g$ being increased, the final 
variance begins to be suppressed, since resonance becomes less efficient
when the $g^2\langle\chi^2\rangle$ term in Eq.~(\ref{phit}) grows 
to the order of $m^2$.  
Numerical simulations based on the Hartree approximation
imply that the final variance slowly 
decreases for $g~\gsim~10^{-4}$.  In order to estimate the 
correct size of fluctuations, 
however, we need to include the rescattering effect 
(i.e., mode-mode coupling), which becomes important when particles are 
sufficiently excited \cite{KT}.  

If the $\chi$ mass is taken into account, 
this generally works as a suppression 
for an efficient preheating.  However, as long as the mass effect does not 
make the $\chi$ field deviate from instability bands, 
 massive $\chi$ particles can be resonantly amplified.  For example 
$\chi$ particles with mass 
$m_{\chi}=m$ are created for $g~\gsim~2 \times 10^{-5}$; and when 
$m_{\chi}=10m$, $g~\gsim~2 \times 10^{-3}$ (See the inset of Fig.~2).  
Since heavy particles whose masses are a few times greater than the 
inflaton mass are sufficiently produced even when  
$g~\gsim~10^{-4}$, this may play an important role for the success
of the GUT scale baryogenesis scenario\cite{baryo1,baryo2}.  It will be 
particularly interesting to estimate the baryon asymmetry produced in
preheating, taking into account the decay of $\chi$ particles as 
in Ref.~\cite{baryo2}.

The spin 1/2 (and higher) quantum fields can  be also produced
during preheating\cite{fermion1,fermion2}, 
although the  Pauli's exclusion principle restricts the 
occupation number as $n_k \le 1$.
Consider a massive fermion field $\psi$ with a bare mass $m_{\psi}$, which is 
coupled to inflaton with a Yukawa interaction, $h\phi\bar{\psi}\psi$.  Then 
the Dirac equation is written in the form: 
\beqn 
(i\gamma^{\mu}\nabla_{\mu}-m_{\rm eff})\psi=0,
\label{dirac}
\eeqn 
where $m_{\rm eff}$ is the effective mass of fermion, defined by
\beqn 
m_{\rm eff}=m_{\psi}+h\phi.
\label{effmass}
\eeqn 
In an expanding universe with the FRW background, fermion production 
takes place around the region where $m_{\rm eff}$ vanishes, 
which corresponds to the value, $\phi_*=-m_{\psi}/h$ \cite{fermion2}. 
Fermions are periodically created when 
the inflaton passes through $\phi_*$, as long 
as the relation $m_{\psi}<|h\Phi|$ is satisfied.  Substituting 
Eq.~(\ref{aapprox}) with Eq.~(\ref{phiapprox}) for Eq.~(\ref{effmass}), 
one finds that the minimum effective 
mass is achieved for $mt=3\pi/2$, yielding the upper limit of massive 
fermion production as 
\beqn 
m_{\psi}~\lsim~h\left(\frac{M_5}{m}\right)^{3/2}m \simeq (10^6h)m.
\label{uplimit}
\eeqn 
This implies that heavy fermions much greater than the inflaton mass are
easily excited unless $h$ is small as $h~\lsim~10^{-6}$.
Since the amplitude of  inflaton decreases slowly relative
the standard preheating, this may delay the time at which fermion  
creations are frozen. 
It will be worth investigating how massive fermions produced in
brane preheating affect the leptogenesis scenario\cite{lepto}.

%%%%%%%%%%%%%%%%%%%%%%%%%%%%%%%%%
 \section{Conclusions and discussions}
%%%%%%%%%%%%%%%%%%%%%%%%%%%%%%%%%

We have studied preheating in the 3-brane world in the massive 
chaotic inflationary scenario. While the existence of the brane in higher 
dimensions works to increase 
the amount of inflation via quadratic modifications
of energy density, its effect in preheating is to make the adiabatic 
damping of the inflaton slower.
Considering the $\chi$ field coupled to inflaton via interaction, $\frac12 
g^2\phi^2\chi^2$, we find that massless $\chi$ particles exhibit strong 
amplification even when the coupling $g$ is small as $g=5 \times 10^{-6}$, 
due to the modifications of the background evolution.  We have also 
verified that massive scalar and fermionic particles heavier than inflaton 
can be nonperturbatively produced, which may play important roles for the 
baryo- and lepto-genesis scenarios.

Although we have restricted ourselves in the massive inflaton case,
the self coupling inflationary scenario 
with an effective potential, $V(\phi)=\frac14 
\lambda\phi^4+\frac12 g^2\phi^2\chi^2$, 
has a different structure of  resonance.
In this case introducing a new variable, $\delta X_k=a\delta\chi_k$, 
Eq.~(\ref{deltachi}) is reduced to the following Lam\'e equation
in the linear stage of preheating:
\beqn 
\frac{d^2}{dx^2}\delta X_k+\left[\kappa^2+\frac{g^2}{\lambda}
{\rm cn}^2\left(x, \frac{1}{\sqrt{2}}\right) \right] \delta X_k=0,
\label{lame}
\eeqn 
where $x\equiv \sqrt{\lambda}\phi_i\eta$ 
and  $\kappa^2 \equiv k^2/(\lambda\phi_i^2)$,
with $\eta \equiv  \int a^{-1}dt$ being a conformal time.
The ${\rm cn}\left(x, 1/\sqrt{2}\right)$ is the elliptic function which 
is well approximated as ${\rm cn}\left(x, 1/\sqrt{2}\right) \approx 
\cos(0.8472x)$.
Since the effect of the adiabatic damping of inflaton does not appear in the 
frequency of  $\delta X_k$, 
the strength of resonance is determined by the ratio, $g^2/\lambda$,
which is the same as the standard scenario discussed in Ref.~\cite{selfpre}. 
Despite this, the slow growth of the scale factor makes the redshift of 
$\delta\chi_k$ less efficient due to the relation of 
$\delta\chi_k=a^{-1}\delta X_k$,
which will provide the larger growth of the variance, $\langle\delta\chi^2\rangle$.

The enhancement of field perturbations can lead to the growth of 
metric perturbations\cite{mpre1,mpre2}, which is particularly important 
for the formation of primordial black holes\cite{PBH}.
This requires a consistent study of cosmological perturbations 
including the effect of the 5-dimensional bulk, 
which we leave to future work.

%%%%%%%%%%%%%%%%%%%%%%%%%%%%%%%%%%%%%%%%%%%%%%%
\section*{ACKNOWLEDGMENTS} 
We thank Bruce A.  Bassett and David Wands 
for useful discussions.  This work was  partially
by the Grant-in-Aid for Scientific Research Fund of the Ministry of
Education, Science and Culture (Specially Promoted Research No.08102010) 
and by the Yamada foundation.

%%%%%%%%%%%%%%%%%%%%%%%%%%%%%%%%%%%%%%%

\end{document}